\documentclass[%
 reprint,
 amsmath,amssymb,
aps,
prl,
]{revtex4-2}

\usepackage{color}
\usepackage{graphicx}% Include figure files
\usepackage{dcolumn}% Align table columns on decimal point
\usepackage{bm}% bold math
\usepackage{comment}
\usepackage{hyperref}% add hypertext capabilities
\begin{document}

\preprint{APS/123-QED}

\title{Extensively parallelizable chiral fermion}% Force line breaks with \\

\author{Yuki Nagai}%
\email{nagai.yuki@jaea.go.jp}
\affiliation{
CCSE, Japan  Atomic Energy Agency, 178-4-4, Wakashiba, Kashiwa, 277-0871, Chiba, Japan
}
\affiliation{
Mathematical Science Team, RIKEN Center for Advanced Intelligence Project (AIP), 1-4-1 Nihonbashi, Chuo-ku, 103-0027, Tokyo, Japan
}

\author{Akio Tomiya}
\email{akio@yukawa.kyoto-u.ac.jp}
\affiliation{
Faculty of Technology and Science, International Professional University of Technology, 3-3-1, Umeda, Kita-ku, Osaka, 530-0001, Osaka,
Japan}%



\date{\today}% It is always \today, today,

\begin{abstract}
Chiral symmetry is a key to investigating quantum physics, from condensed matter to particle physics. We propose a novel way of realizing a chiral fermion, known as the overlap-Dirac operator, without explicitly calculating the low modes of the Wilson-Dirac operator.
We introduce a projection operator inspired by the Sakurai-Sugiura method and formulate the exact sign function and overlap-Dirac operator with a contour-integral form. 
Like the Sakurai-Sugiura method, the proposing method is multi-scale parallelizable, which fits the multi-core/multi-GPGPU paradigm.
We confirm that the quality of chiral symmetry realized with the proposed method is sufficient for double precision. We evaluate the strong scaling of the proposing method.
\end{abstract}

\maketitle

%\tableofcontents

\newcommand{\sgn}{\operatorname{sgn}}

\section{Introduction}
Chiral symmetry in physics plays a crucial role in a wide energy scale. For example, in the standard model of particle physics, chiral symmetry guarantees the lightness of light quarks and pions \cite{Aoki:2021kgd}. In the context beyond the standard model physics, chiral symmetry in supersymmetric Higgs fields protects it from quadratic divergence \cite{Martin:1997ns}. Moreover, it is essential to build the super string theory \cite{Giddings:2001yu}.
In fields of condensed matter physics, the chiral symmetry is important for realizing low-energy spectrum in Graphene or topologically nontrivial materials \cite{Semenoff:2011jf}. 
Nielsen and Ninomiya \cite{Nielsen:1980rz} have shown that chiral symmetry is tightly connected to the discretizations of spacetime and is non-trivial to realize on the lattice. Through the theorem, the structure of chiral symmetry on the lattice is related to the topological insulators  \cite{Witten:2015aoa}.

Lattice gauge theory is a tool to investigate strongly coupled gauge systems like QCD \cite{Karsch:2001cy}. Lattice QCD has been investigated for more than 40 years \cite{Wilson:1974sk}, and recently it is a mandatory technique not only for theoretical physics but also for experimental physics \cite{Zyla:2020zbs}.
The formulation is based on discretized spacetime kept gauge symmetry and evaluated using supercomputers, which achieves great success \cite{PACS-CS:2008bkb, MILC:2009mpl,
RBC:2018dos, Borsanyi:2020mff, Borsanyi:2010cj, HotQCD:2014kol}.
Lattice QCD has demanded a formulation that keeps (precise) chiral symmetry, but that took a long time \cite{Chandrasekharan:2004dph}.

Neuberger introduced the overlap Dirac operator \cite{Neuberger:1997fp},
\begin{align}
D^{\rm ov}(m_q)= \frac{1+m_q}{2}\bm{  1}  + \frac{1-m_q}2 \gamma_5 \sgn(H_W).
\end{align}
where $H_W = \gamma_5 D_W(-m_0)$ is the hermitian Wilson-Dirac operator with negative mass $(-m_0<0)$ and $m_q$ is a fermion mass.
This Dirac-operator has exact chiral symmetry on the lattice \cite{Ginsparg:1981bj,Luscher:1998pqa,Hasenfratz:1997ft} at $m_q =0$.
$\sgn(x)$ is the sign function takes $1$ for $x>0$, $-1$ for $x<0$ and otherwise 0. 
However, the sign function causes difficulty with numerical computations. We must calculate lower-lying eigenvalues and eigenvectors for the Wilson-Dirac operator involved in the overlap-Dirac operator in practice \cite{
vandenEshof:2002ms,JLQCD:2007ppn,Frommer:2011aa}.
In terms of the eigenmodes of $H_W$, the sign function is represented as,
\begin{align}
{\rm sgn}(H_W) 
&=
\sum_i {\rm sgn}(\lambda_i) \vec{v}_i \vec{v}_i^\dagger,
\end{align}
where $\lambda_i$ is an eigenvalue of $H_W$ and $\vec{v}_i$ is the eigenvector for $\lambda_i$. 
Using these arguments, we can evaluate the multiplication of the overlap-Dirac operator on a vector with the use of the Zorotarev approximation \cite{
vandenEshof:2002ms,JLQCD:2007ppn,Frommer:2011aa}. 
\begin{figure}[t]
\includegraphics[width = 7.5cm]{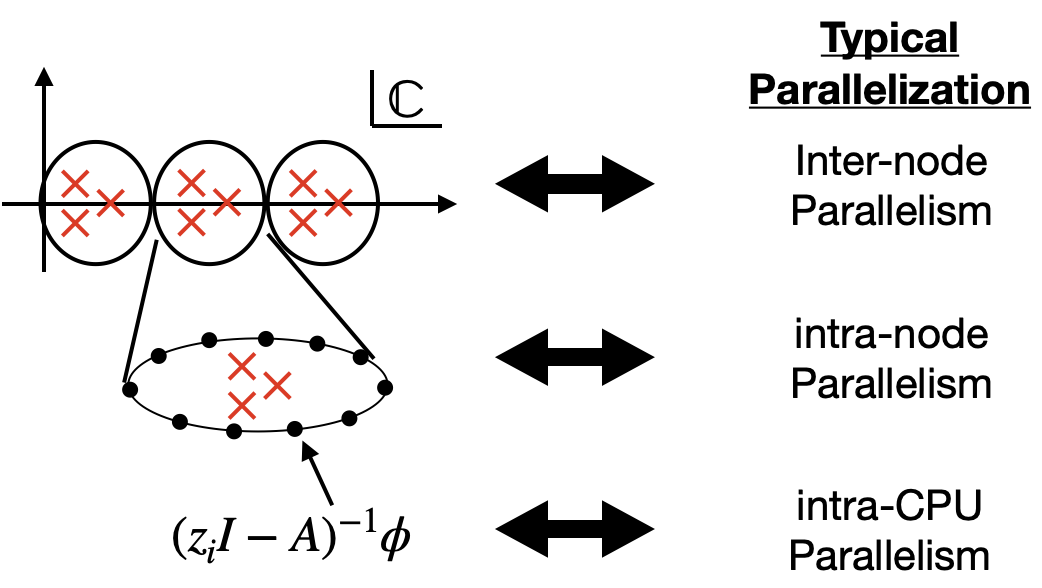}
\caption{\label{fig:ss_para} Schematic figure of multi-level parallelization.
To evaluate the contour integral, it can be parallelized in 3 levels efficiently.
}
\end{figure}
Even using many cores, it cannot be efficiently accelerated because calculations for each core are numerically expensive.

Although the overlap-Dirac operator has numerical difficulty, it has been used in a lot of studies with lattice QCD \cite{Hashimoto:2006rb, Chiu:2008kt, xQCD:2010pnl,
Yagi:2011jn, Cossu:2013uua,
XQCD:2013odc}.
In addition to numerical calculations, mathematical physics {\it e.g.} proof of the index theorem \cite{Fujikawa:1999ku}.
Recent numerical studies with lattice QCD employ the domain-wall fermions \cite{Boyle:2007fn, Chiu:2011dz, Chiu:2013wwa, Chen:2014hva, Shintani:2015vsx, Nakayama:2016atf, Tomiya:2016jwr, Burger:2018fvb, Colangelo:2022jxc, Colquhoun:2022atw, Aoki:2020noz}, which can be regarded as an approximated overlap fermion \cite{Edwards:2000qv}; however,
the approximation occasionally causes pathological behavior \cite{Cossu:2015kfa}.

\begin{figure}[h]
\includegraphics[width = 7.25cm]{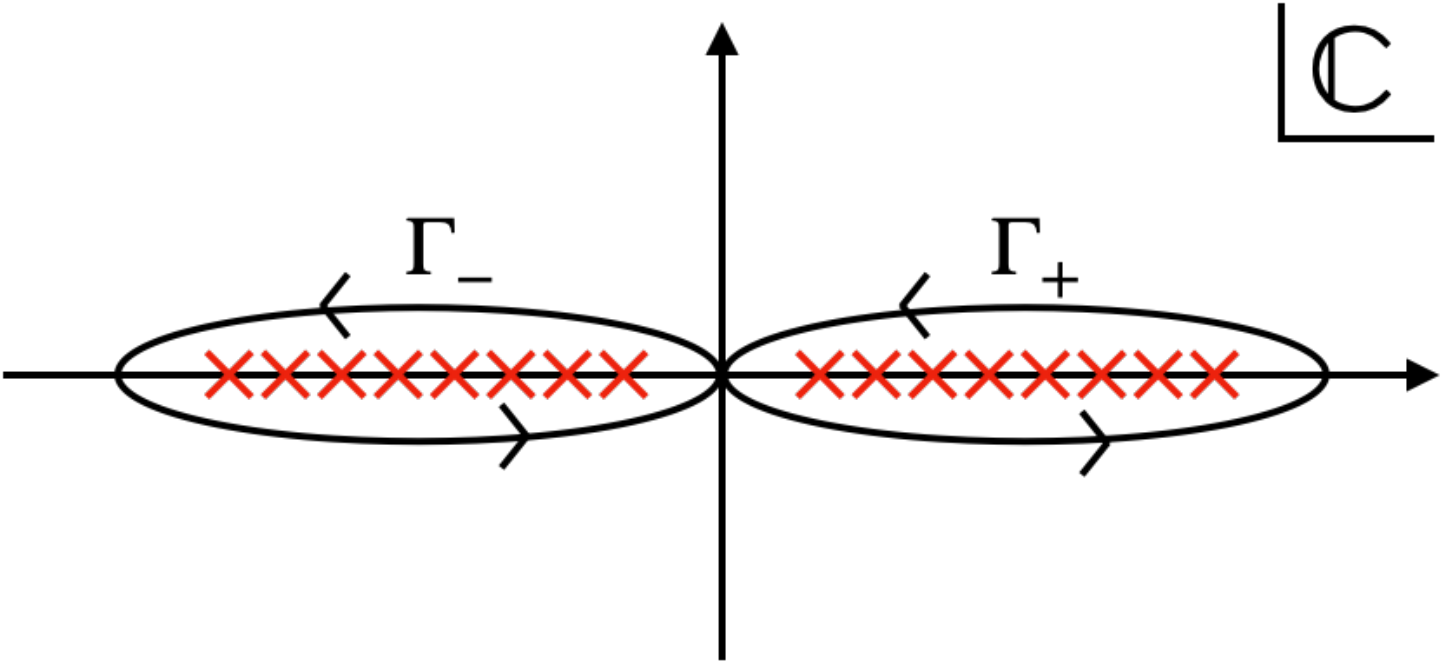}
\caption{\label{fig:complex_plane} 
Schematic figure for the contours to express the sign function $\sgn(H_W)$. Cross symbols represent eigenvalues for $H_W$ except for the zero-modes, which are not contribute the sign function as we described in the main text. Numerical integration on the contours can be parallelized extensively.}
\end{figure}

We propose a completely different and {\it exact} approach to evaluate the matrix-valued sign function. 
We point out that the sign function is expressed as 
\begin{align}
    {\rm sgn}(H_W) = P_+(H_W) - P_-(H_W),
\end{align}
where $P_{+(-)}(H_W) \equiv \sum_{i\in \{i|\lambda_i>(<)0\}} \vec{v}_i \vec{v}_i^{\dagger}$ is a projection operator matrix which projects the matrix $H_W$ to the sub-space spanned by the eigenvectors associated with positive (negative) eigenvalues. 
We emphasize that the zero eigenvalus ($\lambda_i = 0$) do not contribute to the sign function because ${\rm sgn}(\lambda_i)= 0$ by definition of the sign function. 
The projection matrix can be expressed by the contour integral on complex plane \cite{SAKURAI2003119,Nagai_2013}: 
\begin{align}
    P_{\pm}(H_W) = P_{\Gamma_{\pm}}(H_W) \equiv \oint_{\Gamma_{\pm}} \frac{dz}{2\pi i} \frac{1}{z I - H_W},
    \label{eq:projection_in_contour}
\end{align}
since the resolvent of $H_W$ is $(z I - H_W)^{-1} = \sum_{i}  \vec{v}_i \vec{v}_i^{\dagger} /(z - \lambda_i)$. 
Here, all positive (negative) eigenvalues are located inside the closed loop $\Gamma_{+(-)}$ as shown in Fig.~\ref{fig:complex_plane}. 
Therefore, the multiplication of the sign function ${\rm sgn}(H_W)$ and a vector $\vec{x}$ is evaluated by 
\begin{align}
{\rm sgn}(H_W) \vec{x} =  \oint_{\Gamma_{+}} \frac{dz}{2\pi i} \vec{y}(z,H_W,\vec{x})  -  \oint_{\Gamma_{-}} \frac{dz}{2\pi i} \vec{y}(z,H_W,\vec{x}),
\end{align}
where $\vec{y}(z,H_W,\vec{x})$ is a solution of the linear equations
\begin{align}
    (z I - H_W) \vec{y} = \vec{x}. \label{eq:linear}
\end{align}
We remark that we do not have to solve the eigenvalue equation for low-laying modes in contrast to the conventional method \cite{JLQCD:2007ppn}.
% decompose the Dirac operator into two parts, low-lying modes and others, as the conventional method 

We show a method to evaluate a contour integral with numerical quadrature. 
Using $N_q$-point quadrature rule, the vector $ P_{\Gamma} \vec{x}$ is approximately written as 
\begin{align}
     P_{\Gamma} \vec{x} \approx \frac{1}{N_q} \sum_{j=1}^{N_q} \rho w_j \vec{y}_j, \label{eq:integral_decomp}
\end{align}
with $w_j = \alpha \cos \theta_j + i \sin \theta_j$, $z_j = \gamma + \rho (\cos \theta_j + i \alpha \sin \theta_j)$, and $\theta_j = 2\pi(j-1/2)/N_q$. Here, $\alpha$ is a vertical scaling factor. 
The vector $\vec{y}_j =  \vec{y}(z_j,H_W,\vec{x})$ is the solution of the linear equation (\ref{eq:linear}).
We note that, the approximately equal symbol in \eqref{eq:integral_decomp} can be exact in double precision if the $N_q$ is enough large.
This integration can be improve by employing more sophisticated integration scheme.

The linear equations with different $z_j$ (\ref{eq:linear}) can be simultaneously evaluated by a shift solver \cite{Clark:2006wq}.  
In addition, this expression can be extensively parallelized since the integration can be decomposed in small parts. 
We remark that the Sakurai-Sugiura method, which is the eigenvalue solver using the contour integrals, is applicable with multi-scale parallel computations. 
Our proposing method also has the same properties as the parallelization (Fig.~\ref{fig:ss_para}). 
With the use of the shift solver, the calculation of the sign function ${\rm sgn}(H_W) \vec{x}$ consists of two parts: a calculation of a solution at a given seed point $z_{\rm seed}$ by the biconjugate gradient method and constructions of solutions on contours $\Gamma$. 
Therefore, the computational cost is a sum of the matrix-vector multiplication $H_W \vec{v}$ and construction of solutions $\vec{y}_j$. 
In this letter, we do not parallelize the former for simplicity.

In this letter, we examine the expression with the contour integrals \eqref{eq:projection_in_contour}.

\section{Numerical experiments}
In this proof-of-principle study, we perform experiments for 30 quenched configurations generated with SU(3) Wilson gauge action with $\beta = 5.7$ on $L^4=8^4$ lattice using LatticeQCD.jl \cite{LatticeQCDjl}, a package for simulation of the lattice QCD written in Julia programming language. The configurations are generated with the heatbath algorithm by skipping 10 trajectories. There are three overrelaxations per a heatbath update. We employ the standard overlap operator (Wilson kernel) and  we choose $m_0=1$. 
We confirm that our ensemble contains enough fluctuations of the topological charge as described in Supplemental material. 
Detailed extensive study will be provided in another publication.

It is well known that the massless overlap operator has exact chiral symmetry, namely, it satisfies the Ginsparg-Wilson relation \cite{Ginsparg:1981bj,Luscher:1998pqa,Hasenfratz:1997ft},
\begin{align}
\{D^{\rm ov}(0), \gamma_5\} = D^{\rm ov}(0) \gamma_5 D^{\rm ov}(0),
\end{align}
and it is equivalent to $(\sgn(H_W))^2 = I$ \cite{Chiu:2002eh,Cundy:2010uq}.
We evaluate quality of the sign function \cite{Chiu:2002eh}, \begin{align}
\sigma=\frac{\left|\vec{s}^{\dagger} \vec{s}-\vec{y}^{\dagger} \vec{y}\right|}{\vec{y}^{\dagger} \vec{y}}
\end{align}
where $\vec{s} = \sgn(H_W)\vec{y}$ and $\vec{y}$ is a random vector.
This $\sigma$ should be $0$ if the Ginsparg-Wilson relation is exactly satisfied.
Numerically, it is enough that this satisfies in the double precision.

We numerically perform the contour integral in \eqref{eq:projection_in_contour}.
The number of divisions of the numerical integration for the contour integral is taken as $N_{\rm qtot} = 100, 250, 500$ and $1000$ to establish this formulation.
We parallelize the code with MPI to evaluate the integral and compare the number of processes as $1$ -- $120$. In contrast to the conventional way of parallelization for the lattice QCD, we can freely divide the system along with the contour integral independent to the lattice geometry.
We develop a code based on LatticeDiracOperators.jl \cite{LatticeDiracOperators} written in Julia programming language and perform calculations in the supercomputer HPE SGI8600 in the Japan Atomic Energy Agency (Two Intel Xeon Gold 6242R CPUs (3.1GHz, 20core) per a computational node).
%Yukawa21 in YITP.

\subsection{Results}
Here we show results from numerical experiments.
The statistical error is estimated using the Jackknife method. 

%\textcolor{red}{ WE DESCRIBE THE RESULTS IN HERE. ONE BY ONE.}

\begin{figure}[h]
\includegraphics[width = 8cm]{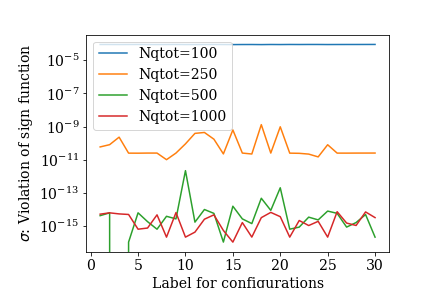}
\caption{\label{fig:sigma_vs_conf} Violation of the sign function as a function of configurations. Different lines correspond to a different number of divisions of the contour integral.}
\end{figure}

First, we show the violation of the sign function for each configuration
(Fig. \ref{fig:sigma_vs_conf}). The violation $\sigma$ does not strongly depend on configurations and instead depends on the number of divisions.
As discussed in the Supplemental material, the result is not correlated to the topology of configurations, which is clear from the definition of the matrix sign function.

\begin{figure}[h]
\includegraphics[width = 8cm]{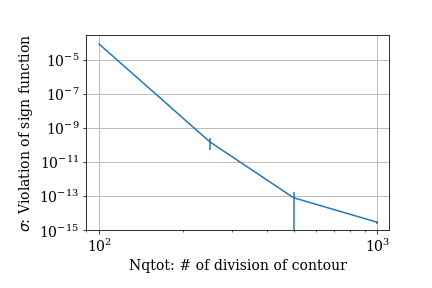}
\caption{\label{fig:sigma_vs_nqtot} Violation of the sign function along with the number of division. $\sigma$ is averaged over configurations. 
The statistical error is evaluated but the same size with the symbol.
}
\end{figure}

%\textcolor{blue}{
Second, we show the result of the violation of the sign function along with the number of divisions (Fig. \ref{fig:sigma_vs_nqtot} ). 
The quality indicator $\sigma$ is averaged over configurations. The error is the same size as the symbol.
One can see that the quality can be improved with the power of the divisions, and this divisions can be parallelly evaluated as in Fig.  \ref{fig:ss_para}.
%}

\begin{figure}[h]
\includegraphics[width = 8cm]{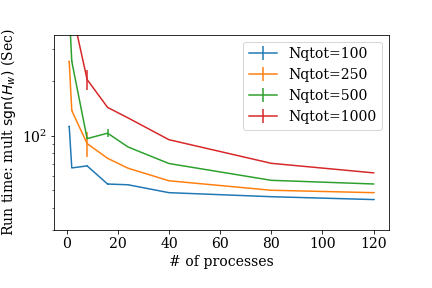}
\caption{\label{fig:time_vs_nprocs} Elapsed time of multiplication $\sgn(H_W)$ to a vector. The time is averaged over configurations. 
The statistical error is evaluated but the same size with the symbol.
}
\end{figure}

%\textcolor{red}{
Fig. \ref{fig:time_vs_nprocs} shows the elapsed time of multiplication $\sgn(H_W)$ to a vector as a functional of the number of the CPU cores. 
The vertical axis is the time that is averaged over configurations.
Each CPU core solves linear equations with different $z_j$, simultaneously, with the use of the shifted Biconjugate Gradient (BICG) method.
The total number of the integration points $N_q$ is divided by the number of the cores.
Since we do not parallelize the matrix-vector multiplication, the elapsed time is governed by the single matrix-vector multiplication when the computational cost to evaluate shifted solutions $\vec{y}_j$ is reduced by many CPU cores (the Amdahl's law). 
The residual elapsed time due to matrix-vector operation can be reduced by the low-level parallelization as shown in Fig.~\ref{fig:ss_para}. 
The most expensive case in our parameter $N_{\rm qtot}=1000$ is shown in Tab. \ref{table:runtimeHW}
It should be noted that the actual value of the elapsed time is not important in this table since we do not use fully optimized code and algorithm. 
For example, there might be better contour shape, position of the seed, and convergence criteria to reduce the computational cost. 

In summary,
$N_{\rm qtot} \gtrsim 500$ achieves enough quality, $\sigma \lesssim 10^{-14}$ and the elapsed time is roughly $70$ seconds with more than $50$ processors.
%We note that, we have not been parallelized along with \textcolor{red}{hoge}, and it should scales much better than the current implementation but we leave the scaling study for the future.

\begin{table}[h]
\caption{Elapsed time for multiplying $\sgn(H_W)$ with $N_{\rm qtot}=1000$}
\label{table:runtimeHW}
 \centering
  \begin{tabular}{ccc}
\hline
\# of procs &
time (sec) &
$\sigma$ \\
\hline\hline
  1 & 959(7) & (2.97 $\pm 0.60$) $\times 10^{-15}$  \\
  2 & 493(3) & (3.0$\pm 0.6$) $\times 10^{-15}$  \\
 16 & 142.8(3) & (2.9 $\pm 0.6$) $\times 10^{-15}$  \\
 24 & 125.0(2) & (3.0 $\pm 0.7$) $\times 10^{-15}$  \\
 40 &  94.8(3) & (2.9 $\pm 0.7$) $\times 10^{-15}$  \\
 80 &  70.3(3) & (2.95 $\pm 0.63$) $\times 10^{-15}$  \\
120 &  62.3(3) & (2.9 $\pm 0.6$) $\times 10^{-15}$  \\
   \hline
  \end{tabular}
\end{table}

\section{Summary}
In this work, we introduce a novel way to realize the overlap-Dirac operator without calculating eigenvalues and eigenvectors explicitly.
We introduce projection operators represented by contour integrals involving the Wilson-Dirac operator $H_W$. 
Similar to the Sakurai-Sugiura method \cite{SAKURAI2003119, Nagai_2013}, our method can be parallelized very efficiently.

The proposed method would pave the way for extensive studies of chiral physics from condensed matter to high energy on many-core/multi-GPGPU supercomputers.

\begin{acknowledgments}
The authors thank Akinori Tanaka for fruitful discussions.
The part of this research was conducted with the supercomputer HPE SGI8600 in the Japan Atomic Energy Agency.  
%Kakenhi. 
Numerical computation in this work was partially carried out at the Yukawa Institute Computer Facility.
This work is partially supported by JSPS KAKENHI Grant Numbers  22K03539 (YN and AT).
\end{acknowledgments}

\bibliography{ref}

%%%%%%%%%% Merge with supplemental materials %%%%%%%%%%
\pagebreak
\widetext
\begin{center}
\textbf{\large Supplemental Materials: Extensively parallelizable chiral fermion}
\end{center}
%%%%%%%%%% Merge with supplemental materials %%%%%%%%%%
%%%%%%%%%% Prefix a "S" to all equations, figures, tables and reset the counter %%%%%%%%%%
\setcounter{equation}{0}
\setcounter{figure}{0}
\setcounter{table}{0}
\setcounter{page}{1}
\makeatletter
\renewcommand{\theequation}{S\arabic{equation}}
\renewcommand{\thefigure}{S\arabic{figure}}
\renewcommand{\bibnumfmt}[1]{[S#1]}
\renewcommand{\citenumfont}[1]{S#1}
%%%%%%%%%% Prefix a "S" to all equations, figures, tables and reset the counter %%%%%%%%%%

%\appendix
\begin{comment}
\section{Evaluation of contour integrals with numerical quadrature}
We show a method to evaluate a contour integral with numerical quadrature. 
The projection operator with a given elliptic domain $\Gamma$ can be calculated by 
\begin{align}
    P_{\Gamma} {\bm v}= \oint_{\Gamma} \frac{dz}{2\pi i} (z I - H_{W})^{-1} {\bm v}.
\end{align}
Using $N_q$-point quadrature rule, the vector $ P_{\Gamma} {\bm v}$ is approximately written as 
\begin{align}
     P_{\Gamma} {\bm v} \approx \frac{1}{N_q} \sum_{j=1}^{N_q} \rho w_j {\bm y}_j, \label{eq:integral_decomp}
\end{align}
with $w_j = \alpha \cos \theta_j + i \sin \theta_j$, $z_j = \gamma + \rho (\cos \theta_j + i \alpha \sin \theta_j)$, and $\theta_j = 2\pi(j-1/2)/N_q$. Here, $\alpha$ is a vertical scaling factor. 
The vector ${\bm y}_j$ is the solution of the linear equation $(z_j I - H_W) {\bm y}_j = {\bm v}$. 
Note that, the approximately equal symbol in \eqref{eq:integral_decomp} can be exact in double precision if the $N_q$ is enough large.
This integration can be improve by employing more sophisticated integration scheme.
\end{comment}

\section{Numerical setup for the integration}
In the contour integral with numerical quadrature, we set a vertical scaling factor $\alpha = 0.1$, a radius $\rho = 1.1999$ and the origin of the ellipse $\gamma = \pm 1.2$. 
We note that we can tune parameters if all eigenvalues are located inside the contours. 
We set the seed point, $z_{\rm seed} = 1+0.05i$, where the linear equation should be solved.
For simplicity, we do not use a seed switching technique \cite{doi:10.1143/JPSJ.77.114713}.

%parameters should be written in here.

\section{Topological sector}
The overlap Dirac operator is tightly related to the topological charge through the index theorem.
In particular, a configuration $Q\neq0$ accompanies zero-modes of $D^{\rm ov}(0)$.

\begin{figure}[h]
\includegraphics[width = 10cm]{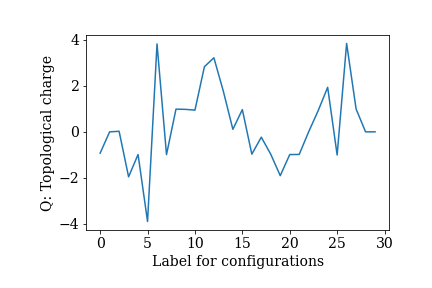}
\caption{\label{fig:Qhist} 
History of the topological charge.
}
\end{figure}

First, we examine the topological fluctuation of our ensemble. 
We employ $O(a^2)$ improved definition of the topological charge and measure it after the Wilson flow $t=4.0$ to make them approximately integers. 
Fig. \ref{fig:Qhist} is the history of the topological charge and it shows that our ensemble contains $Q\neq0$ configurations.

% - - - - - - - - - - - - - - - - - - - - - - - - 

\begin{figure}[h]
\includegraphics[width = 10cm]{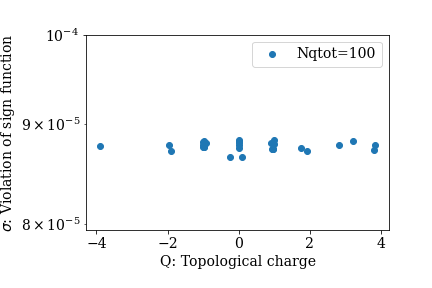}
\caption{\label{fig:sigma_vs_Q_Nqtot100} 
Correlation between the violation of sign function $\sigma$ and topological charge for $N_{\rm qtot}=100$.
}
\end{figure}

\begin{figure}[h]
\includegraphics[width = 10cm]{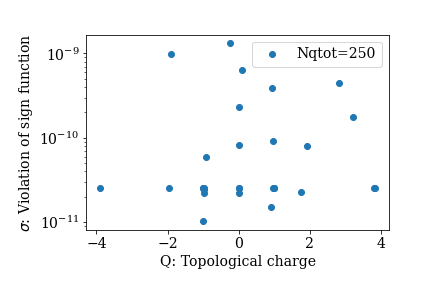}
\caption{\label{fig:sigma_vs_Q_Nqtot250} 
Same figure to Fig. \ref{fig:sigma_vs_Q_Nqtot100} but for $N_{\rm qtot}=250$.
}
\end{figure}

\begin{figure}[h]
\includegraphics[width = 10cm]{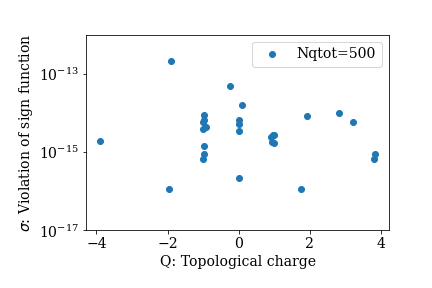}
\caption{\label{fig:sigma_vs_Q_Nqtot500} 
Same figure to Fig. \ref{fig:sigma_vs_Q_Nqtot100} but for $N_{\rm qtot}=500$.
One data point reaches $\sigma=0$ numerically and it is droped from the figure.
}
\end{figure}

\begin{figure}[h]
\includegraphics[width = 10cm]{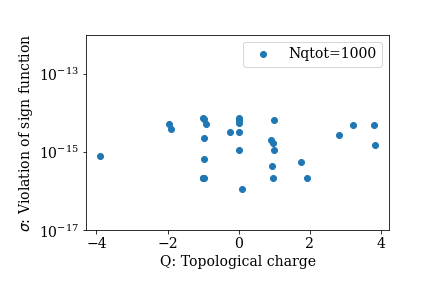}
\caption{\label{fig:sigma_vs_Q_Nqtot1000} 
Same figure to Fig. \ref{fig:sigma_vs_Q_Nqtot100} but for $N_{\rm qtot}=1000$.
}
\end{figure}

%\section{Appendix D: Correlation between topological charge and the quality of the sign function}
Next, we see the correlation between the topological charge and the violation of sign function $\sgn(H_W)$. As we have mentioned, $\sgn(x)=0$ for $x=0$ in the introduction; however, some readers might suspect topological zero-modes could affect the quality, so we check the correlation between them.
Fig. \ref{fig:sigma_vs_Q_Nqtot100} -- \ref{fig:sigma_vs_Q_Nqtot1000} show the correlation for $ N_{\rm qtot}$.
One can see that, no clear correlation between the topological sector and the violation $\sigma$.


\end{document}